\documentclass[10pt,prb,twocolumn,superscriptaddress,citeautoscript]{revtex4-1}
\usepackage{graphicx}
\usepackage{color}
\newcommand{\im}{\rm i}

\begin{document}

\title{Disentangling electronic and vibronic coherences in two-dimensional echo spectra}
\author{Christoph Kreisbeck}
\affiliation{Department of Chemistry and Chemical Biology, Harvard University, Cambridge, Massachusetts 02138, USA}
\author{Tobias Kramer}
\affiliation{Institut f\"ur Physik, Humboldt Universit\"at zu Berlin, 12489 Berlin, Germany}
\affiliation{Department of Physics, Harvard University, Cambridge, Massachusetts 02138, USA}
\author{Al\'an Aspuru-Guzik}
\affiliation{Department of Chemistry and Chemical Biology, Harvard University, Cambridge, Massachusetts 02138, USA}
\date{\today} 

\begin{abstract}
The prevalence of long-lasting oscillatory signals in the 2d echo-spectroscopy of light-harvesting complexes has led to a search for possible mechanisms.
We investigate how two causes of oscillatory signals are intertwined: (i) electronic coherences supporting delocalized wave-like motion, and (ii) narrow bands in the vibronic spectral density. 
To disentangle the vibronic and electronic contributions we introduce a time-windowed Fourier transform of the signal amplitude. 
We find that 2d spectra can be dominated by excitations of pathways which are absent in excitonic energy transport.
This leads to an underestimation of the life-time of electronic coherences by 2d spectra.
\end{abstract}
\maketitle



\section{Introduction}

The observation of long-lasting oscillations in peak amplitudes of two-dimensional (2d) echo spectroscopy as function of increasing delay time between pump and probe pulses 
\cite{Engel2007a,Lee2007a,Collini2010a,Calhoun2009a,Panitchayangkoon2010a,Turner2012a}
has lead to an intensive search for quantum mechanical effects in photosynthetic systems at physiological temperatures.
The general theoretical model of excitonic energy transfer from the antenna to the reaction center is well established \cite{Amerongen2000a,Blankenship2002a,Frigaard2004a}.
For the Fenna-Matthews-Olson (FMO) complex an efficient transfer relies on a dissipative coupling of electronic and vibronic degrees of freedom \cite{Mohseni2008a,Rebentrost2009a,Plenio2008a,Caruso2009a,Wu2010a,Mohseni2011a,Kreisbeck2011a}, which puts the excited system in an energy funnel towards the reaction center.
The oscillations recorded in 2d echo spectra were not anticipated within a simplified rate equation formalism based on Markovian approximations for the dissipative coupling of vibronic and electronic degrees of freedom \cite{Cho2005a,Brixner2005a}.

\begin{figure}
\begin{center}
\includegraphics[width=0.9\columnwidth]{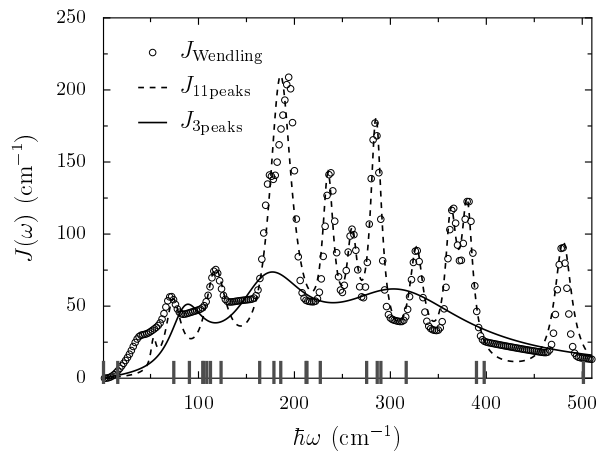}
\end{center}
\caption{\label{fig:FMOspectraldensity}
Spectral density of the FMO complex. Circles: measured spectral density \cite{Wendling2000a}, parametrization \cite{Renger2006a}.
Fits $J_{\rm \{3,11\} peaks}$ are used for the GPU-HEOM calculation. The marks on the frequency axis indicate differences of exciton eigenenergies.
}
\end{figure}

The physical origin of the experimental observations has lead to various theoretical proposals, which are roughly divided into two mechanisms. As shown from calculations in model Hamiltonians \cite{Christensson2012a,Chin2013a,Kolli2012a} the presence of $\delta$-spiked peaks in the vibronic mode distribution does imprint a long-lasting oscillatory signal on coherences. 
The spectral density determined from fluorescence line-narrowing experiments ($J_{\rm Wendling}$ in Fig.~\ref{fig:FMOspectraldensity}) reveals a series of peaks. 
However a calculation of 2d spectra and coherences \cite{Kreisbeck2012a} using a close approximation to the density (denoted by $J_{\rm 3 peaks}$ and $J_{\rm 11 peaks}$ in Fig.~\ref{fig:FMOspectraldensity}) puts the vibronic contributions on a much smaller scale compared to electronic coherences, which lead to oscillation periods determined by the differences in excitonic eigenenergies.
The second proposed mechanism \cite{Kreisbeck2012a} identifies two prerequisites for long-lasting electronic coherences and oscillations of cross-peaks of 2d-spectra in the continuous part of the vibronic mode distribution: (i) a small initial slope of the vibronic density towards zero frequency, and (ii) a larger coupling between a continuum of vibrations whose frequencies are in the range of the typical excitonic energy gaps. 
Both conditions are fullfilled by the florescence line-narrowing spectral density $J_{\rm Wendling}$.

It remains an open question to what extend 2d spectra are probing vibrational or electronic properties of light-harvesting complexes. 
Moreover, for the design of artificial nanostructure it is of interest to identify the physical mechanism for efficient transport and conditions for the accompanying vibronic density. 
Here, we disentangle the interplay of electronic and vibronic effects on the transport by introducing a reliable method to 
dissect the 2d-echo signal. 
We show that a time-windowed short-time Fourier transform (STFT) is a suitable tool to track and assign specific vibronic and electronic contributions at all delay times. 
We find that two-dimensional echo-spectra can significantly {\em underestimate} the duration of electronic coherences due to the destructive superposition of several pathways related to additional excitations and ground-state bleaching. 
Such pathways are not important for excitonic energy transfer, which for the same system displays longer electronic coherences than the 2d spectra.
The connection between 2d spectra and efficient transport and the role of coherences requires a careful evaluation.

\section{Transition from electronic to vibrational induced coherence in the FMO complex}

\begin{figure}
\begin{center}
\includegraphics[width=0.9\columnwidth]{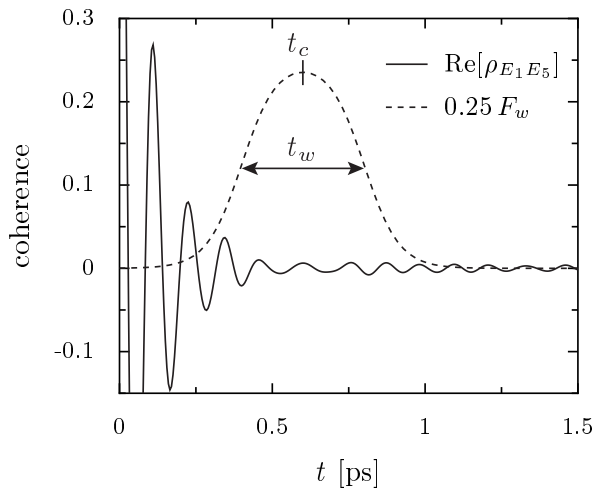}
\end{center}
\caption{\label{Plot_WendlingFMO_E1E5_illustarteWindowFunction}
Time evolution of the coherence $\rho_{E_1E_5}(t)$ for the spectral density $J_{\rm 11peaks}$ at $T=150$~K. The dashed line shows a typical window function ($t_c=0.6$~ps, $t_w=0.4$~ps, $\beta=7$).
}
\end{figure}

\begin{figure}
\begin{center}
\includegraphics[width=0.9\columnwidth]{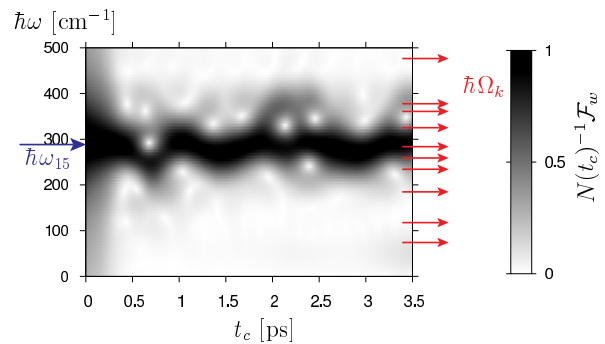}
\end{center}
\caption{\label{Plot_Polish_WendlingFMO_Crossover_tw0_8ps_Elec_Vib_E1E5}
STFT analysis ($t_w=0.4$~ps, $\beta=7$) of the coherence $\rho_{E_1E_5}$ of Fig.~\ref{Plot_WendlingFMO_E1E5_illustarteWindowFunction}.
The left arrow marks the frequency of the electronic energy difference $E_1-E_5$ and the right arrows
indicate the centers of vibrational peaks of $J_{\rm 11 peaks}$, Fig.~\ref{fig:FMOspectraldensity}.
}
\end{figure}

Multiple frequencies are a common pattern in the cross-peak oscillations observed in 2d echo spectra of
various LHCs such as the FMO complex \cite{Hayes2011a, Panitchayangkoon2011a} or light harvesting
proteins PE545 and PC645 from marine cryptophyte algae \cite{Collini2010a,Turner2012a}.
To unravel which physical processes are present as the delay-time progresses and to uncover the transition from electronic to vibrational induced coherences we introduce a windowed short-time Fourier transform (STFT) of the signal
\begin{equation}\label{eq:11}
\mathcal{F}_w(\omega,t_c)\;\rho_{E_iE_k}(t)=\int_{-\infty}^\infty \mbox{d}t\, F_{w}(t_c,t)
\rho_{E_iE_k}(t)e^{\im \omega t}
\end{equation}
where $t_c$ denotes the center and $t_w$ the width of the window function
\begin{equation}\label{eq:12}
F_{w}(t_c,t)=\sum_{s=\pm 1}\frac{s}{1+e^{\beta(t-t_c-s t_w/2)/t_w}}. 
\end{equation}
Later, we normalize $\mathcal{F}_w(t_c)\rho_{E_iE_k}$ to its maximal value. 
A typical window function along with the coherence between exciton eigenstates $|E_1\rangle$ and $|E_5\rangle$ of the FMO complex calculated in Ref.~\cite{Kreisbeck2012a} is shown in Fig.~\ref{Plot_WendlingFMO_E1E5_illustarteWindowFunction} together with the corresponding STFT analysis in Fig.~\ref{Plot_Polish_WendlingFMO_Crossover_tw0_8ps_Elec_Vib_E1E5}.
Sweeping the centre of the window function $t_c$ yields information about the changes in the frequency distribution with advancing time in the exciton dynamics.
In Fig.~\ref{Plot_Polish_WendlingFMO_Crossover_tw0_8ps_Elec_Vib_E1E5}, the initially dominant electronic coherent frequency 
$\hbar\omega_{15}$ is indicated by the arrow on the left side.
At later times dephasing reduces the influence of the electronic coherence while the interaction with the strongly coupled vibrational modes increases. 
Around $t_c=0.7$~ps a ``gap'' can be seen, where the electronic frequency is absent but shortly afterwards re-emerges.
From $1-3.5$~ps a wobbling motion of the dominant contribution is seen, which is caused by the vibronic frequencies marked by arrows on the right hand side. 
During the time evolution, energy is shared among different vibrational modes
close in resonance with the exction frequency $\hbar\omega_{15}$. Which vibrations are activated
changes with time. For example, the vibration at $\hbar\Omega=380$~cm$^{-1}$ is present around $2.2$~ps but absent around $3$~ps.
The features revealed in the STFT reflect two different mechanisms in the energy transfer.
Electronic coherence signifies delocalized wavelike energy transfer which helps to overcome energy barriers and facilitates fast energy transfer to specific target states. 
On the other hand the signatures of strongly coupled vibrational modes mark reversible energy exchange between the exciton system and the protein environment \cite{Rebentrost2011a}. 

\section{Coherences in 2d spectra}

\begin{figure}
\begin{center}
\includegraphics[width=0.9\columnwidth]{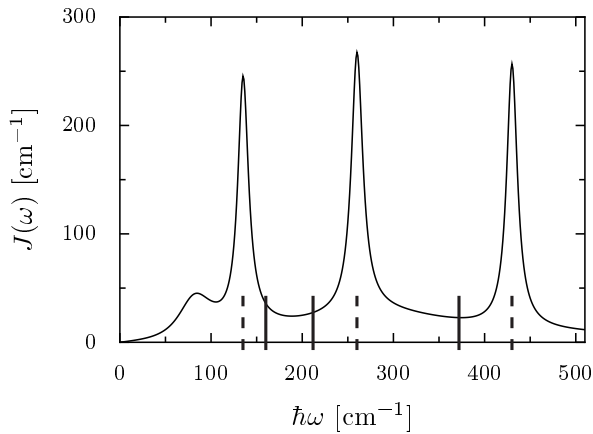}
\caption{\label{Plot_SpectralDensity_All}
Five peak model spectral density $J_{\rm 5 peaks}$ with a suppressed slope at zero frequency and three strongly coupled
underdamped vibrational modes. The vibrational frequencies (dashed line) are close in resonance with
the difference of the exciton energies $E_i-E_j$ (solid line) of the exciton system Eq.~(\ref{eq:4}).
Parameters for Eq.~(\ref{eq:specdens}): $\lambda_k\in\{11,9,12,7,4\}$~cm$^{-1}$, 
$\nu_k^{-1}\in\{230,50,750,700,750\}$~fs,  $\hbar\Omega_k\in\{80,260,135,260,430\}$~cm$^{-1}$.
}
\end{center}
\end{figure}

\begin{figure}
\begin{center}
\includegraphics[width=0.9\columnwidth]{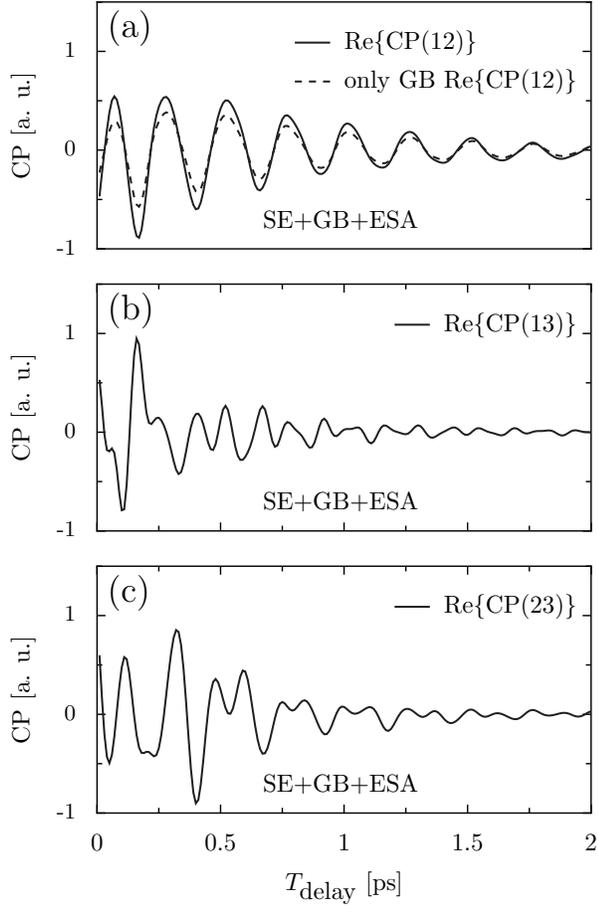}
\end{center}
\caption{\label{Plot_CrossPeaks_SEGBESA}
Oscillatory component of the real part of the cross-peak dynamics of the total rephasing
2d-echo signal for Eq.~(\ref{eq:4}) and $J_{\rm 5 peaks}$ as function of delay time $T_{\rm delay}$. 
The temperature is set to $T=150$~K.
}
\end{figure}

\begin{figure}
\begin{center}
\includegraphics[width=0.9\columnwidth]{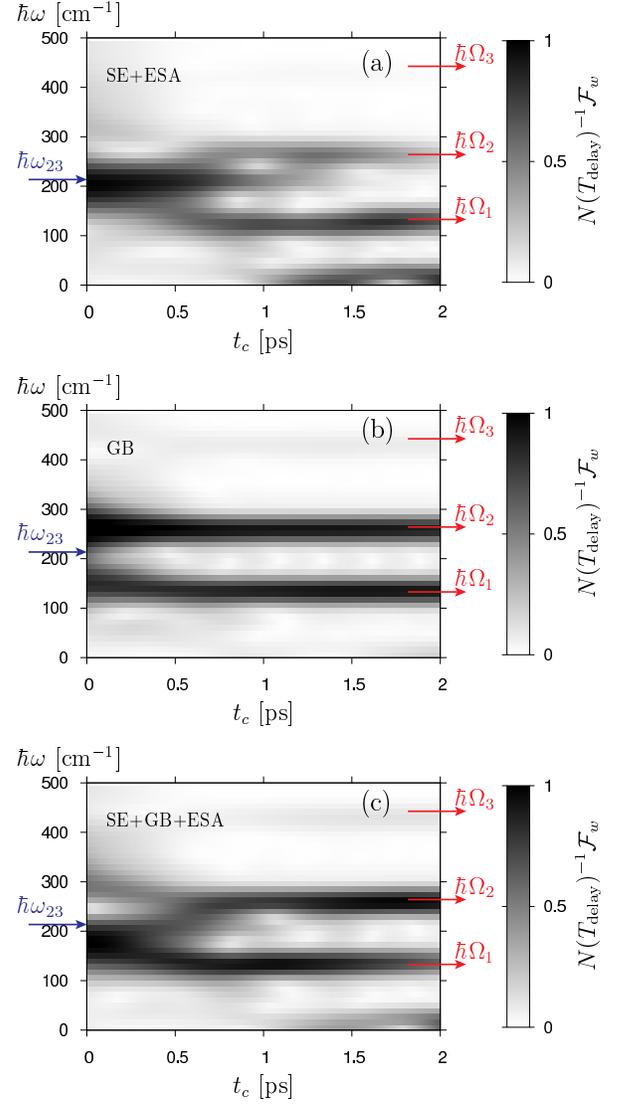}
\end{center}
\caption{\label{Plot_Crossover_tw700fs_beta7_Elec_Vib_Peak23}
STFT analysis ($t_w=0.7$~ps, $\beta=7$) of the cross-peak dynamics CP(23) of the rephasing 2d-echo signal.
Shown are results for (a) the sum of the SE and ESA pathway, (b) the contribution of the GB pathway and
(c) the total signal including all three pathways.
The width of the window function is $t_w=1.0$~ps and $T=150$~K. The left arrow marks the frequency
of electronic coherence $\hbar\omega_{23}$ and the right arrows mark the frequencies of the strongly
coupled vibrational modes $\hbar\Omega_k$.
}
\end{figure}

The decay of peak amplitude oscillations in two-dimensional spectra has been used to extract coherence lifetimes \cite{Panitchayangkoon2010a}. 
In the following, we identify two processes which reduce the signature of electronic coherence in two-dimensional spectra compared to transport calculations.
To elucidate and quantify the interplay between electronic and vibronic degrees of freedom we discuss the physical processes mapped by 2d-echo spectra within a model trimer consisting of sites $1-3$ of the FMO complex \cite{Renger2006a}
\begin{equation}\label{eq:4}
 \mathcal{H}_{\rm ex}=
\left(\begin{array}{ccc}
 410 & -87.7 & 5.5 \\  
 -87.7 & 530 & 30.8 \\ 
  5.5 & 30.8 & 210  
\end{array}\right)
\mbox{cm}^{-1}.
\end{equation}
As for the FMO complex \cite{Kreisbeck2012a} we describe the energy transfer within a Frenkel exciton model \cite{May2004a}, linearly coupled to independent baths at each site.
The spectral density is parametrized by a superposition of shifted Drude-Lorentz peaks
\begin{equation} \label{eq:specdens}
 J(\omega)=\sum_{k=1}^{M}\left[\frac{\nu_k\lambda_k\omega}{\nu_k^2+(\omega+\Omega_k)^2}+\frac{\nu_k\lambda_k\omega}{\nu_k^2+(\omega-\Omega_k)^2}\right].
\end{equation}
The spectral density $J_{\rm 5peaks}$ (Fig.~\ref{Plot_SpectralDensity_All}) for the model trimer fulfils the criteria for long-lived electronic coherences due to its small initial slope and later on larger coupling.
To study the role of specific vibrational modes $J_{\rm 5 peaks}$ adds three vibrational modes centered at frequencies $\hbar\Omega_1=135$~cm$^{-1}$, $\hbar\Omega_2=260$~cm$^{-1}$, $\hbar\Omega_3=430$~cm$^{-1}$, which puts them close to resonance with the differences in exciton eigenenergies.
The total 2d-echo rephasing signal is composed of three Liouville pathways \cite{Mukamel1999a,Cho2009a} reflecting stimulated emission (SE), ground-state bleaching (GB) and excited state absorption (ESA), where the laser pulses create two excitons in the system.
Formally, we calculate the Fourier transform of the third order response function $S_{RP}(t_3,T_{\rm delay},t_1)$ \cite{Mukamel1999a,Cho2009a} in the impulsive limit
\begin{eqnarray}\label{eq:13}
I_{RP}(\omega_1,T_d,\omega_3)
&=&\int_0^{\infty}\int_0^{\infty}\mbox{d}t_1\mbox{d} t_3 e^{\im\omega_3t_3-\im\omega_1t_1}\\\nonumber
&&\!\!\!\!\!\!\!\!\!\!\!\!\!\!\!\!\!\!\!\!\!\!\!\!\!\!\!\!\!\!\!\!\!\!\!\!\!\!\times[S_{RP}^{GB}(t_3,T_d,t_1)+S_{RP}^{SE}(t_3,T_d,t_1)+S_{RP}^{ESA}(t_3,T_d,t_1)]
\end{eqnarray}
We proceed along the steps in Ref.~\cite{Hein2012a} and include the rotational average over random orientations by sampling 20 distinct laser polarization vectors aligned along the vertices of a dodecahedron. 
We assume equal dipole strength of the three pigments and dipole orientations along the nitrogen atoms $N_B-N_D$ \cite{Adolphs2008a}.
The propagation is performed with the GPU-HEOM method \cite{Kreisbeck2011a,Kreisbeck2013a}.
In an experimental setup all three pathways are tied together and cannot be analysed separately, while the theoretical calculations allows us to study the individual contributions.
Typically, the SE and ESA pathways interfere destructively and lead to a diminished amplitude of the
coherent signal in the total spectrum.
The cancellation of coherent amplitudes in 2d spectra enhances the influence of strongly coupled vibrations on the cross-peak oscillations.
In particular the ground-state vibrations monitored by the GB get amplified in weight in the 2d-echo spectra. 
Our finding, that ground-state vibrations affect the cross-peak dynamics is in agreement with Tiwari $et$ $al.$ \cite{Tiwari2012a} who investigate the effect of a single vibrational mode in resonance with a model dimer.
Different conditions govern transport and 2d-echo spectra, where several pulses hit the sample.
For instance the ground-state bleaching and excited state absorption induced by the 2d-echo setup are not relevant for the energy transfer process.
Thus the apparent diminishing of coherences in 2d spectra does not imply unimportance of coherence in the excitonic transfer.\\
The cancellation effects are prominently visible in Fig.~\ref{Plot_CrossPeaks_SEGBESA}, which displays the oscillatory component of the lower-diagonal cross-peak dynamics of CP(12), CP(13) and CP(23) of the rephasing signal as function of delay time $T_{\rm delay}$. 
The amplitude of CP($ik$) is determined by integrating the real part of the rephasing 2d-spectra (all three rephasing pathways) over a small rectangle ($\Delta E=36$~cm$^{-1}$) in the 2d-energy grid.
The center of the rectangular area is located at $\omega_i=E_i+\lambda/4.5$ and $\omega_k=E_k+\lambda/4.5$, where $E_i, E_k$ denote the exciton eigenenergies of Eq.~(\ref{eq:4}).
The shift $\lambda/4.5$ takes into account the stokes shift and ensures that the amplitude is averaged around the maximum of the peaks. 
The non-oscillatory background is fitted by $f(t)=a+b\,e^{-ct} + d\, t + d_2\, t^2$ and is subtracted from the total signal.
The three cross peaks show different behaviours.
In the upper panel CP(12) oscillates dominantly with a single frequency that matches the vibrational mode $\hbar\Omega_1$. 
The 2d-spectra thus apparently shows that there is almost no contribution of electronic coherence
between exciton states $|E_1\rangle$ and $|E_2\rangle$ left. 
This finding is in contradiction to simulations of the coherence $\rho_{E_1E_2}(t)$ that predict long lasting electronic coherence up to $1$~ps. 
Studying the dynamics of the SE and ESA pathways of CP(12) separately recovers the long lasting electronic coherence. 
But in the case of CP(12) the amplitudes of the coherent oscillations of the SE and ESA pathway are nearly identical and
the destructive interference of these pathways cancels the contribution of  electronic coherence in the total signal. 
The remaining oscillatory component of the total signal reflects the ground state vibrations of the GB pathway shown as dashed line in  Fig.~\ref{Plot_CrossPeaks_SEGBESA}(a) and the signal of electronic coherence is masked by the vibrational mode $\hbar\Omega_1$.
The cross-peaks CP(13) and CP(23) (Fig.~\ref{Plot_CrossPeaks_SEGBESA}(b) and Fig.~\ref{Plot_CrossPeaks_SEGBESA}(c)) show a more complex dynamics, which is best analysed using the STFT methodology for the case of CP(23) in Fig.~\ref{Plot_Crossover_tw700fs_beta7_Elec_Vib_Peak23}.
Multiple frequencies originating from electronic coherence and vibronic modes superpose each other and form rich structures in
the beating pattern. 
The STFT frequency analysis for the sum of ESA and SE pathways is depicted in Fig.~\ref{Plot_Crossover_tw700fs_beta7_Elec_Vib_Peak23}(a).
In agreement with the coherences (not shown) we observe a transition around $T_{\rm delay}=0.7$~ps from initially electronic coherence with
frequency $\hbar\omega_{23}=212$~cm$^{-1}$ to vibrational frequencies $\hbar\Omega_{k}$ at later times. 
The strongest coupling to the exciton dynamics stems from the vibrational modes $\hbar\Omega_1$ and $\hbar\Omega_2$
that are close in resonance with the electronic frequency $\hbar\omega_{23}$. 
For the GB pathway (Fig.~\ref{Plot_Crossover_tw700fs_beta7_Elec_Vib_Peak23}(b)) the exciton system remains in the electronic ground state during $T_{\rm delay}$ and no electronic frequencies show up in the cross peak dynamics.
In the total signal Fig.~\ref{Plot_Crossover_tw700fs_beta7_Elec_Vib_Peak23}(c) the GB vibrational modes leave their trace and diminish the relative weight of the electronic coherences.

\section{Conclusion}

The STFT provides a tool to determine the physical mechanisms behind oscillatory signals for progressing delay time. 
We find that the determination of electronic coherence life-times from the experimentally observed total signal in the 2d-echo spectra is difficult and can underestimate the lifetime of electronic coherences.
For varying delay times the STFT shows multiple frequencies and a mix of vibronic and electronic contributions. 
Similar observations are reported for experimentally recorded 2d spectra \cite{Hayes2011a, Panitchayangkoon2011a,Collini2010a,Turner2012a}.
We further found examples of a reemergent amplitude in the oscillatory component at later time in CP(23), where a larger amplitude at $T_{\rm delay}=0.35$~ps is seen compared to $T_{\rm delay}=0.2$~ps.
Interestingly, such reemerging amplitudes are also observed for the FMO complex \cite{Hayes2011a, Panitchayangkoon2011a}. 

The pollution of the 2d-signal by ground-state vibrations and two-exciton states is not directly relevant for assessing the role of coherences in excitonic energy transfer, where these two pathways are absent. 
This has also implications for identifying vibronic spectral-densities supporting efficient transport.
We find a good correspondence between coherence lifetimes of the stimulated emission pathway (SE) of cross peak CP$(i,j)$ and the corresponding coherence $\rho_{E_iE_j}$ .
For certain cross-peaks the ground state vibrations dominate, and the 2d-echo spectra are not able to determine the  electronic coherence lifetime. 
In this case an additional analysis, like the proposed witness of electronic coherence \cite{Yuen2012a} or quantum process tomography \cite{Yuen2011a,Yuen2011b} is required to reveal the hidden information. 

\acknowledgements
A.A.-G. and C.K. are supported by the DARPA grants N66001-10-1-4063, N66001-10-1-4059, and T.K. by a Heisenberg fellowship of the DFG (KR~2889/5). A.A.-G. thanks the Corning foundation for their generous support.

\providecommand*\mcitethebibliography{\thebibliography}
\csname @ifundefined\endcsname{endmcitethebibliography}
  {\let\endmcitethebibliography\endthebibliography}{}

\end{document}